\documentstyle[aps,preprint]{revtex}
\begin{document}
\draft
\title{Conductance Fluctuations in a Metallic Wire \\
Interrupted by a Tunnel Junction}
\author{A. van Oudenaarden, M. H. Devoret$^*$, E. H. Visscher, Yu. V. Nazarov,
and J. E. Mooij}
\address{Department of Applied Physics and \\
Delft Institute of \\
Micro-electronics and Submicron-technology (DIMES), \\
Delft University of Technology, PO Box 5046, 2600 GA Delft, The Netherlands}
\date{\today }
\maketitle

\begin{abstract}
The conductance fluctuations of a metallic wire which is interrupted by a
small tunnel junction has been explored experimentally.
In this system, the bias voltage $V$, which drops almost completely inside
the tunnel barrier, is used to probe the energy dependence of conductance
fluctuations due to disorder in the wire.
We find that the variance of the fluctuations is directly
proportional to $V$. The experimental data are consistently described by a
theoretical model with two phenomenological parameters: the phase breaking
time at low temperatures and the diffusion coefficient.
\end{abstract}

\pacs{PACS numbers: 73.40.Rw, 73.23.-b, 72.15.Gd}

\vspace{0.5cm}

Mesoscopic quantum interference phenomena in metallic wires, like weak
localization and universal conductance fluctuations, are manifestations of
the wave nature of electrons which are robust to microscopic disorder \cite
{review1,review2}.
Changing the microscopic disorder of a mesoscopic metallic wire
results in a change
of the conductance of the order of the universal value $e^2/h$.
Experimentally one generally measures the conductance fluctuations
resulting from a change in the magnetic field, rather than from a change in
the microscopic disorder.
A large collection of experimental data from metallic to poorly conducting
systems convincingly confirms the universal nature of the conductance
fluctuations \cite{webb}.
Although the influence of microscopic disorder
is well understood, much
less is known of the influence of the electron energy on the conductance
fluctuations. In conventional experiments \cite{webb}
on metallic wires it is very difficult
to change the energy of the electrons in a controlled way. However when the
wire is interrupted by a tunnel junction
it is possible to systematically
examine the conductance fluctuations as a function of electron energy.
The bias voltage over the wire drops almost completely inside
the tunnel barrier,
because the resistance of the tunnel junction is much larger than the
resistance of the wire.
The junction serves then as an injector of electrons in a specific
energy range which can be tuned by the bias voltage.
In this Letter we report on measurements
of the conductance fluctuations in a metallic wire interrupted by a
tunnel junction. We
explored the dependence of the variance of the fluctuations on the bias
voltage, magnetic field and temperature. In particular we measured
autocorrelation functions to determine the correlation field and correlation
voltage. We compare the full set of experimental data directly to
theoretical predictions \cite{yuli}.

The fluctuations of the differential
conductance across the wire+junction are due to interference of electron
waves that return to the tunnel barrier after tunneling. In Fig.~1a such a
trajectory is depicted. When an electron tunnels through the junction at
point {\bf a} it diffuses through the right electrode and returns after
numerous scatter events at point {\bf b}. After tunneling to the left
electrode at {\bf b} it diffuses back to point {\bf a}. The dashed line in
Fig.~1a denotes the trajectory of another electron wave which energy differs
by $\varepsilon $ from the energy of the trajectory depicted by the full
line. If the dephasing time $\tau _\varphi $ is smaller than $\hbar
/\varepsilon $, the two partial waves propagate coherently and
constructively interfere.
An analogous interference effect occurs in a
normal metal-superconductor tunnel junction. Here the subgap conductivity
is determined by the
interference of electron and hole waves returning to the junction
\cite{nstheory,nsexperimental}.
The maximum length of the trajectory in Fig.~1a is $\Lambda
=v_F\tau _\varphi $, where $v_F$ is the Fermi velocity. The dominant
contributions to the conductance fluctuations are those caused by the type
of trajectories depicted in Fig.~1a with length $\Lambda $\cite{yuli}. The
diffusing electrons can coherently penetrate in the electrode over a
distance $L_\varphi =\sqrt{D\tau _\varphi }$ (Fig.~1a), where $D$ is the
diffusion coefficient.
The finite dephasing time $\tau _\varphi $ results in an energy broadening
of $\gamma =\hbar /\tau _\varphi $. The normalized variance of the
conductance fluctuations is given by\cite{yuli}:
\begin{equation}
\frac{\left\langle \delta G^2\right\rangle }{G_T^2}\approx \frac{eV}{\gamma
^3}\Delta _L\Delta _R,
\end{equation}
where $\Delta _L$ and $\Delta _R$ are the typical level spacing in the left
and right electrode respectively and $G_T$ is the tunneling conductance. The
angle brackets denote an average over impurity configurations. The former
equation can be rewritten in terms of the quantum conductance $G_Q\equiv
\frac{e^2}\hbar $ and the effective resistances $R_L^\varphi $ and
$R_R^\varphi $ of the left and right lead \cite{thouless}:
\begin{equation}
\frac{\left\langle \delta G^2\right\rangle }{G_T^2}\approx \frac{eV}\gamma
G_Q^2R_L^\varphi R_R^\varphi .
\end{equation}
The resistances $R_L^\varphi $ and $R_R^\varphi $ are formed by a part of
the lead with length $L_\varphi $ (Fig. 1b). In contrast with conductance
fluctuations of a single wire the conductance fluctuations due to the leads
of a single tunnel junction are non-universal. The variance $\left\langle
\delta G^2\right\rangle $ is directly proportional to $\frac{eV}\gamma $,
which is the number of energy slices with width $\gamma $ above the Fermi
energy. Each energy slice fluctuates independently. When the bias voltage is
increased, the number of fluctuating slices increases linearly.
The presence of the tunnel junction opens the possibility to change
the number of independently fluctuating slices in a controlled way.
This is very difficult in a single metallic wire.
Even
though the resistances of the leads are typically 10$^4$ times smaller than
the tunnel resistance, the fluctuations due to the leads can be of the order
of 1\% of the total resistance. This phenomenon totally contradicts
classical addition formulas for resistances and is a pure quantum interference
effect. In order to accurately measure the conductance fluctuations the
single junction has to be surrounded by a well defined low impedance
environment. In this case the zero-bias anomaly is almost suppressed \cite
{ingold} and the fluctuations can be measured on a well characterized
background.

The sample was fabricated using electron beam lithography and a multilayer
process \cite{visscher}. A schematic layout of the sample is shown in
Fig.~1b. A small Al-Al$_2$O$_3$-Al tunnel junction was fabricated using a
shadow evaporation technique. The junction capacitance, estimated from the
overlap area of the junction, is 2 fF. The tunnel conductance $G_T$ is 57
$\mu $S. The small junction is shunted on chip by a capacitance $C_S$ and a
resistance $R_S$. The shunt capacitor is a large parallel plate capacitor of
200$\times $200 $\mu $m$^2$ with 75 nm SiO as dielectric. The shunt
capacitance, measured at room temperature, is approximately 45 pF. The shunt
resistor is provided by a platinum strip between the capacitor plates. The
resistance $R_S$, determined by a four terminal measurement at low
temperatures, is 28 $\Omega .$ The left and right junction leads have a
width $W_L$ and $W_R$ of 200 nm and 400 nm, a thickness $t_L$ and $t_R$ of
25 nm and 40 nm.

The measurements were performed in a dilution refrigerator. The measurement
leads were filtered by means of RC and copper powder microwave filters at
the temperature of the mixing chamber. At room temperature the leads were
additionally filtered by feedthrough $\Pi $-filters. The sample was mounted
in a microwave tight 'box-in-a-box' construction at the temperature of the
mixing chamber.

The differential conductance $G$ of the tunnel junction was measured with a
lock-in technique. In Fig.~2a and 2b $G$ is plotted as a function of the
bias voltage $V$. In Fig. 2c $G$ is shown versus the magnetic field $B$.
Note that $B$ is large enough to drive the aluminium into the normal state.
The conductance fluctuations both as a function of $V$ and $B$ are
reproduced when the experiment is repeated. The
raw-data
differential conductance
has a small minimum at zero voltage bias (not shown in Fig. 2a). This
zero-bias anomaly is due to inelastic tunnel events \cite{ingold}. In our
sample this zero-bias anomaly is controlled by the low impedance environment
resistor $R_S$. To eliminate the effect of the zero-bias anomaly from the
conductance fluctuations, we subtracted from
the raw-data differential conductance
a term proportional to
$V^{R_SG_Q/\pi }$. The result is shown in Fig.~2a and 2b. The conductance
fluctuations grow with increasing bias voltage. Changing the magnetic field
is similar to changing the impurity configuration in the leads (ergodic
hypothesis). By sweeping $B$ we get an effective ensemble average. We
measured the variance $\left\langle \delta G^2\right\rangle $ using a data
set of 5000 points, in which $G$ is measured from 0.2 T to 1.7 T at a fixed
bias voltage $V$. In Fig. 3 $\left\langle \delta G^2\right\rangle $
normalized to the square of the tunnel conductance $G_T$ is shown as a
function of $V$ at different temperatures. For $V<$ 0.8 mV the normalized
variance is directly proportional to $V$ as predicted by (2). At larger $V$
a saturation is observed probably due to out of equilibrium effects. In the
inset of Fig.~3 the temperature dependence at $V$ = 0.8 mV is depicted.

The trajectory of a diffusing electron which returns to the tunnel barrier
encloses, assuming Brownian motion, a typical area $\left\langle
S^2\right\rangle ^{1/2}=\frac 1{\sqrt{12}}L_\varphi W$, where $W$ denotes
the electrode width (Fig. 1a). When the magnetic field perpendicular to the
electrode plane is changed by $\Phi _o/\left\langle S^2\right\rangle ^{1/2}$
the two electron waves cannot longer interfere constructively. Here $\Phi _o$
is the flux quantum $h/e$. A quantitative analysis can be done by extracting
autocorrelation functions. In Fig.~4 the magnetoconductance correlation
function $\left\langle \delta G(B)\delta G(B+\Delta B)\right\rangle $
normalized to the variance is plotted as a function of $\Delta B$. The
correlation function is computed from a set of 5000 data points collected
from 0.2 T to 1.7 T at $V$ = 0.8 mV and $T$ = 20 mK. In the inset of Fig.~4
the correlation function $\left\langle \delta G(V)\delta G(V+\Delta
V)\right\rangle $ as a function of $\Delta V$ is plotted. This correlation
function is calculated using 3000 data points from $V=-0.8$ mV to $V=0.8$
mV. We eliminated the effect of the zero-bias anomaly as described before.
The full expression for the correlation functions is given by: 
\begin{equation}
\frac{\left\langle \delta G(V,B)\delta G(V+\Delta V,B+\Delta B)\right\rangle 
}{G_T^2}=\frac{4V}{\pi e\hbar \nu _L\nu _RV_m^2}\int_0^\infty dt\text{ }%
P_{cl}^LP_{cl}^R\cos (\frac e\hbar \Delta Vt)e^{-2t/\tau _\varphi
}e^{-t/\tau _B}J_1^2(\frac e\hbar V_mt)
\end{equation}
Here $\nu _L$ and $\nu _R$ are the density of states at the Fermi level in
the left and right electrodes. The classical probability function
$P_{cl}^{L,R}(t)$ expresses the chance that an electron that leaves the
barrier on the left or right side returns to the barrier in a time $t$.
Assuming one-dimensional diffusion ($W\ll L_\varphi $),
$P_{cl}^{L,R}(t)=(S_{L,R}\sqrt{\pi D_{L,R}t})^{-1}$, where $D_L$ and $D_R$
are the diffusion coefficients in the left and right electrode and $S_L$ and 
$S_R$ the cross sections of the left and right electrode.

The times $\tau _\varphi $ and $\tau _B$ describe the loss of coherence due
to inelastic processes and due to the magnetic field respectively. The
dephasing time due to the magnetic field is given by $\tau _B=12\left( \frac
\hbar {e\Delta B}\right) ^2\frac 1{D_LW_L^2+D_RW_R^2}$\cite{beenakker}. The
factor 2 in the exponent of equation (3) is due to the fact that both the
left and right electrode contribute a term $e^{-t/\tau _\varphi }$ to the
integral. Another mechanism which provides a long-time cut-off is the finite
instrumental resolution of the lock-in. The amplitude of the modulation
voltage $V_m$ of the lock-in was 3 $\mu V$. In equation (3) this
instrumental dephasing is governed by the square of the first order Bessel
function of the first kind $J_1^2(\frac e\hbar V_mt)$.

The dephasing time $\tau _\varphi $ can be determined directly from the
correlation function $\left\langle \delta G(V,B)\delta G(V+\Delta
V,B)\right\rangle $. Using (3) we obtain $\tau _\varphi $ = 289 $\pm $ 10
ps at $T$ = 20 mK, which corresponds to an energy broadening $\gamma $ of
2.3 $\mu $eV. The fit result is shown as the full line in the inset of Fig.
4. The dephasing cannot be due to the Nyquist noise of the shunt resistor
$R_S$. Nyquist noise would cause a dephasing time
$\tau _N=\frac{\hbar /e^2}{R_S}\frac \hbar {k_BT}$
of 50 ns at $T$ = 20 mK, which is more than two
orders larger than $\tau _\varphi $. We therefore introduce a
phenomenological dephasing time $\tau _\varphi (0)$, which accounts for
dephasing at the lowest temperatures.
At larger temperatures the effective dephasing time is then given by
$\tau _\varphi ^{-1}(T)=\tau _\varphi ^{-1}(0)+\tau_N^{-1}(T).$
From $\left\langle \delta G(V,B)\delta
G(V,B+\Delta B)\right\rangle $ we can extract the diffusion coefficient $D_L$
= 0.0039$\pm $0.0002 m$^2$/s and $D_R$ = 0.0063$\pm $0.0002 m$^2$/s
(full line in Fig. 4), which are compatible with previous experiments \cite
{nsexperimental}. Here we assumed that the ratio of the thicknesses $t_L/t_R$
equals $D_L/D_R.$ Using these data we find a coherence length $L_\varphi $
of 1.2 $\mu m$ at $T=20$ mK. The experimental values of the diffusion
coefficient are consistent with $D=\frac 13v_F\ell $, where $\ell $ is the
mean free path. Considering the thickness of the electrodes, $\ell $ is of
the order of 10 nm. When we use the free electron Fermi velocity
$v_F=2\times 10^6$ m/s \cite{kittel}, we obtain $D\approx 0.007$ m$^2$/s. The
normalized correlation functions are determined with an accuracy of
approximately 5 \%. We calculate the variance at low temperatures using (3)
and a density of states $\nu _L=\nu _R=2.2\times 10^{47}$ J$^{-1}$m$^{-3}$,
which is obtained from specific heat experiments \cite{kittel}. For $T$ = 20
mK and $V$ = 0.8 mV we obtain $\left\langle \delta G^2\right\rangle
/G_T^2=8\times 10^{-6}$. This value is consistent with the experimental
value $\left\langle \delta G^2\right\rangle /G_T^2=1\times 10^{-5}$. We
obtained the same results for another sample with $G_T=62$ $\mu $S. The
variance at larger temperatures agrees with the predictions of (3) when
Nyquist noise by a shunt resistor $R_S$ of 55 $\Omega $ is taking into
account (full line in the inset of Fig.~3).
For $T>$ 1 K the experimental values are smaller than the
theoretical prediction assuming Nyquist noise. Probably at high temperatures
other dephasing mechanisms start to play a role (e.g. electron-electron and
electron-phonon interaction).

In summary, we have measured the conductance fluctuations in a diffusive wire
which is interrupted by a tunnel barrier. The full set of experimental data
can consistently be described by two phenomenological parameters, the
residual phase breaking time and the diffusion coefficient. This last
parameter is consistent with estimates of the mean free path. This type of
experiments opens a new way of measuring the
energy-dependent
properties of diffusing
electrons in disordered media.

We thank the Delft Institute of Micro-electronics and Submicron-technology
(DIMES) for the support in fabrication. The financial support of the Dutch
Foundation for Fundamental Research on Matter (FOM) is acknowledged.

$^{*}$ Permanent address: Service de Physique de l'Etat Condens\'{e},
CEA-Saclay, F-91191, Gif-sur-Yvette, France.

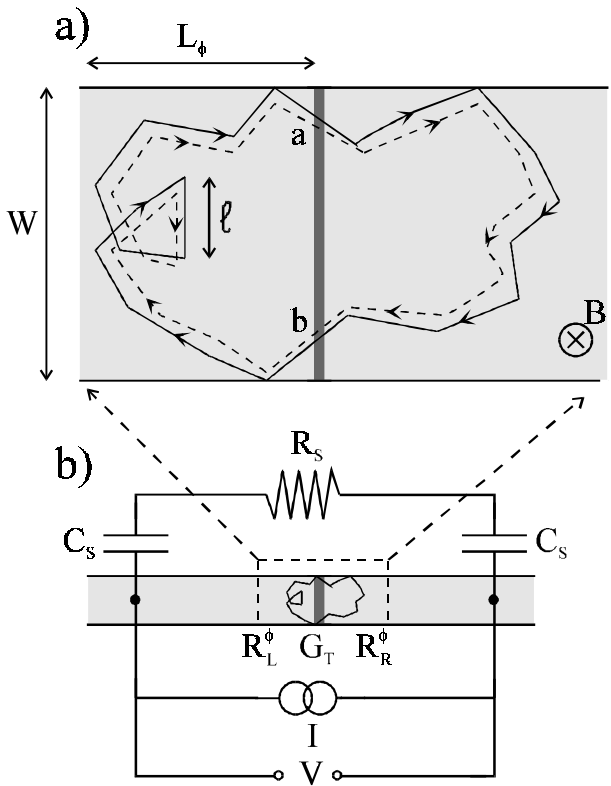
\begin{figure}[tbp]
\caption{a) Class of trajectories which dominantly contribute to the
conductance fluctuations. The direction of the magnetic field is
perpendicular to the junction plane. b) Electromagnetic environement of the
diffusive wire and the tunnel junction. The light grey area denotes the
diffusive wire and the dark grey area the tunnel barrier.}
\end{figure}

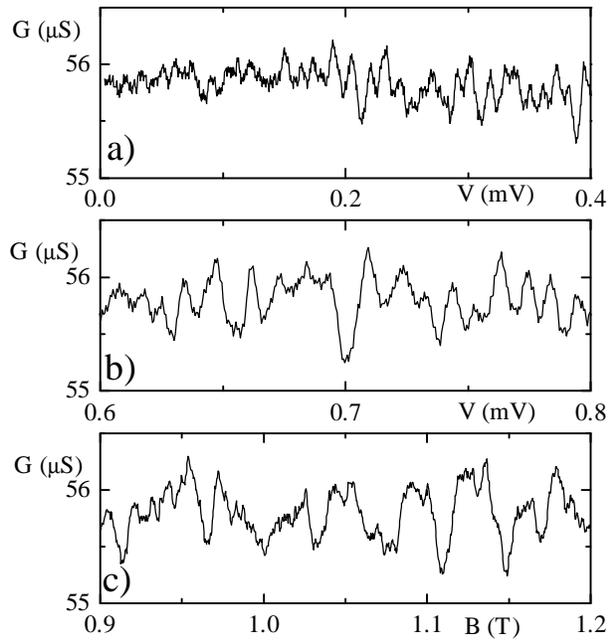
\begin{figure}[tbp]
\caption{a,b) Differential conductance $G$ as a function of bias voltage
$V$ at $T$ = 20 mK and $B$ = 0.5 T; c) The differential conductance $G$ as a
function of magnetic field $B$ at $T$ = 20 mK and $V$ = 0.8 mV.}
\end{figure}

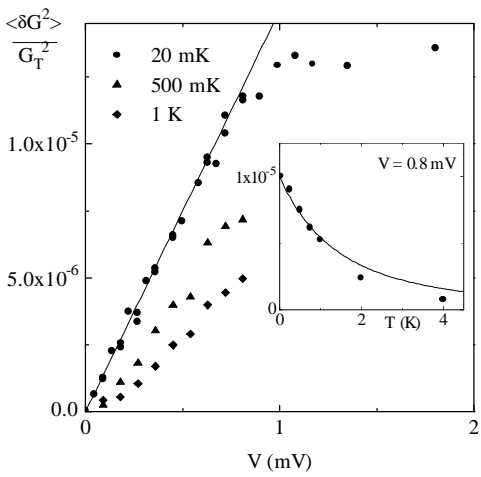
\begin{figure}[tbp]
\caption{Normalized variance as a function of $V$
for different temperatures. In the inset the
normalized variance as a function of $T$ is shown at $V$ = 0.8 mV.
The full line in the inset
denotes the theoretical result when Nyquist noise of
the shunt resistor is the source of dephasing.}
\end{figure}

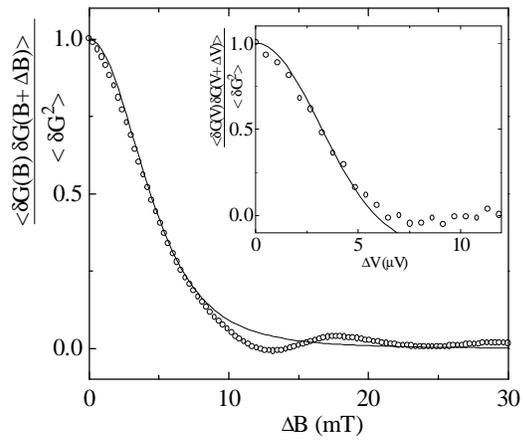
\begin{figure}[tbp]
\caption{Magnetoconductance correlation function at $T$ = 20 mK and $V$ =
0.8 mV. The full line denotes the theoretical result with
$D_L$ = 0.0039 m$^2 $/s
and $D_R$ = 0.0063 m$^2$/s. In the inset the voltage correlation
function is shown for $T$ = 20 mK and $B$ = 0.5 T. The full line represents
the theoretical result for $\tau_\varphi$ = 289 ps.}
\end{figure}

\end{document}